\documentclass{llncs}
\usepackage{lmodern}
\usepackage[T1]{fontenc}
\usepackage{amsmath}
\usepackage{amssymb}
\usepackage{stmaryrd}
\usepackage{graphicx}
\usepackage{xcolor}
\usepackage{hyperref}
\usepackage{xurl}
\usepackage{listings}
\usepackage{isabelle-listings}
\usepackage{comment}
\usepackage[color]{zed}

\definecolor{codeblue}{rgb}{0.1,0.1,0.7}
\definecolor{codegray}{rgb}{0.5,0.5,0.5}
\definecolor{codepurple}{rgb}{0.58,0,0.82}
\definecolor{backcolour}{rgb}{0.95,0.95,0.92}

\lstdefinestyle{pythonstyle}{
    backgroundcolor=\color{backcolour},
    commentstyle=\color{codegray},
    keywordstyle=\color{codeblue}\bfseries,
    stringstyle=\color{codepurple},
    basicstyle=\ttfamily\footnotesize,
    breakatwhitespace=false,
    breaklines=true,
    captionpos=b,
    keepspaces=true,
    numbers=left,
    numbersep=5pt,
    showspaces=false,
    showstringspaces=false,
    showtabs=false,
    tabsize=4
}

\lstdefinelanguage{Isabelle2}{
  morekeywords={theory, imports, begin, lemma, proof, show, have, qed, by, simp, text},
  sensitive=true,
  morecomment=[l]{--},        
  morecomment=[s]{(*}{*)},    
  morestring=[b]",
}

\lstdefinestyle{isabellestyle}{
    backgroundcolor=\color{backcolour},
    basicstyle=\ttfamily\footnotesize,
    breakatwhitespace=false,
    breaklines=true,
    captionpos=b,
    keepspaces=true,
    numbers=left,
    numbersep=5pt,
    showspaces=false,
    showstringspaces=false,
    showtabs=false,
    tabsize=2
  }

\DeclareUnicodeCharacter{00AC}{$\neg$}
\DeclareUnicodeCharacter{03B1}{$\alpha$}
\DeclareUnicodeCharacter{03B2}{$\beta$}
\DeclareUnicodeCharacter{03B3}{$\gamma$}
\DeclareUnicodeCharacter{03B4}{$\delta$}
\DeclareUnicodeCharacter{03B5}{$\varepsilon$}
\DeclareUnicodeCharacter{03B7}{$\eta$} 
\DeclareUnicodeCharacter{03B8}{$\theta$} 
\DeclareUnicodeCharacter{03BB}{$\lambda$}
\DeclareUnicodeCharacter{03BC}{$\mu$}
\DeclareUnicodeCharacter{03BE}{$\xi$}
\DeclareUnicodeCharacter{03C3}{$\sigma$} 
\DeclareUnicodeCharacter{03C9}{$\omega$} 
\DeclareUnicodeCharacter{2032}{${}^\prime$}
\DeclareUnicodeCharacter{21A6}{$\mapsto$}
\DeclareUnicodeCharacter{21C0}{$\rightharpoonup$}
\DeclareUnicodeCharacter{21CC}{$\rightleftharpoons$}
\DeclareUnicodeCharacter{21D2}{$\Rightarrow$}
\DeclareUnicodeCharacter{21E9}{$\downarrow$}
\DeclareUnicodeCharacter{21F8}{$\pfun$}
\DeclareUnicodeCharacter{211D}{$\mathbb{R}$}
\DeclareUnicodeCharacter{2115}{$\mathbb{N}$}
\DeclareUnicodeCharacter{2200}{$\forall$}
\DeclareUnicodeCharacter{2203}{$\exists$}
\DeclareUnicodeCharacter{2204}{$\nexists$}
\DeclareUnicodeCharacter{2208}{$\in$}
\DeclareUnicodeCharacter{2209}{$\notin$}
\DeclareUnicodeCharacter{2211}{$\Sigma$} 
\DeclareUnicodeCharacter{2218}{$\circ$}
\DeclareUnicodeCharacter{2219}{$\bullet$}
\DeclareUnicodeCharacter{221E}{$\infty$} 
\DeclareUnicodeCharacter{2225}{$\|$}
\DeclareUnicodeCharacter{2227}{$\wedge$}
\DeclareUnicodeCharacter{2228}{$\vee$}
\DeclareUnicodeCharacter{2229}{$\cap$}
\DeclareUnicodeCharacter{222A}{$\cup$}
\DeclareUnicodeCharacter{2260}{$\neq$}
\DeclareUnicodeCharacter{2261}{$\equiv$}
\DeclareUnicodeCharacter{2264}{$\le$}
\DeclareUnicodeCharacter{2265}{$\ge$}
\DeclareUnicodeCharacter{2286}{$\subseteq$}
\DeclareUnicodeCharacter{2295}{$\oplus$}
\DeclareUnicodeCharacter{22A5}{$\bot$}
\DeclareUnicodeCharacter{22C0}{$\bigwedge$}
\DeclareUnicodeCharacter{2308}{$\lceil$}
\DeclareUnicodeCharacter{2309}{$\rceil$}
\DeclareUnicodeCharacter{27E6}{$\llbracket$}
\DeclareUnicodeCharacter{27E7}{$\rrbracket$}
\DeclareUnicodeCharacter{27F6}{$\longrightarrow$}
\DeclareUnicodeCharacter{27F7}{$\longleftrightarrow$}
\DeclareUnicodeCharacter{27F9}{$\Longrightarrow$}
\DeclareUnicodeCharacter{290F}{$\longrightarrow$}
\DeclareUnicodeCharacter{2987}{$\llparenthesis$}
\DeclareUnicodeCharacter{2988}{$\rrparenthesis$}

\renewcommand{\ttdefault}{txtt}

\newcommand{\isalogo}{\includegraphics[width=9pt]{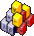}}
\newcommand{\isalink}[1]{\href{#1}{\isalogo}}

\title{Verifying Numerical Methods with Isabelle/HOL}
\author{Dustin Bryant\inst{1} \and Jonathan Julian Huerta y Munive\inst{2} \and Simon Foster\inst{3}}
\institute{Independent, USA \and University of Aalborg, Copenhagen, Denmark \and University of York, UK}

\authorrunning{Bryant, Huerta y Munive, and Foster}

\date{\today}

\begin{document}

\maketitle

\begin{abstract}
Modern machine learning pipelines and ODE solvers are built on numerical algorithms. Reliable numerical methods are thus a prerequisite for trustworthy machine learning and cyber-physical systems. We evaluate a framework designed for verifying imperative programs and the Isabelle proof assistant as tools for proving the total correctness of four numerical algorithms: the bisection method, the fixed-point method, the perceptron, and the gradient descent algorithm. Our verifications required subtle extensions and generalisations to Isabelle's version of Taylor's theorem and higher-order derivatives. Finally, we reflect on the framework's automation, friendly syntax, and on further requirements to turn it into a verification tool for numerical methods.
\end{abstract}

\section{Introduction}
\label{sec:intro}

Numerical methods are an important class of algorithms that iteratively approximate solutions to mathematical problems. They often employ concepts from linear algebra and analysis, and are foundational to machine learning. For instance, stochastic gradient descent, the backbone of modern neural network training, minimises cost functions by approximating their gradients and updating the network's weights accordingly. While such algorithms have well-known mathematical foundations, work on their formal verification is still in its infancy.

Verification of numerical algorithms is challenging for several reasons. Firstly, their formalisation requires mature libraries of diverse mathematical concepts, such as matrices, transcendental functions, and their derivatives. Secondly, despite the existence of various methods for verifying iterative methods~\cite{KellisonA22,Pasca10,Immler12}, their implementations still require expertise in functional programming and in the idiosyncrasies of the verification tools employed. Finally, proof automation has remained a bottleneck for the adoption of these tools.
We therefore investigate whether verifying numerical methods with an interface closer to the best-known imperative programming paradigm, and that also alleviates specific prover knowledge with automation, is feasible. 

If possible, this would contribute to enabling a wider audience to verify numerical methods by lowering the barrier to using these high-assurance tools. Two developments make the evaluation of this question more achievable: the growth of formalised libraries of Analysis and Ordinary Differential Equations~\cite{HolzlIH13,Immler14,ImmlerT16,mathlib2020}, and the creation of frameworks for imperative program verification on top of these~\cite{Foster2020-IsabelleUTP,Foster2024ITrees,HuertaFGSLH2024}. Considering this, we contribute a qualitative assessment of one of these imperative programming frameworks by using it to verify the total correctness of four numerical algorithms.

We choose the Isabelle/ITrees framework~\cite{Foster2024ITrees} that provides user-friendly programming syntax together with understandable Hoare logic specifications and automation for verification condition generation (VCG). Discharging the resulting VCs only requires application and development of foundational theorems from multivariate analysis, many of which are provided by Isabelle's \textsf{HOL-Analysis} package. Isabelle's \textsf{Sledgehammer} tool largely aids in the proving process, and substantially automates the verification. Moreover, the use of ITrees means we can execute the algorithms in Isabelle and generate code that intrinsically has a high degree of assurance in its correctness.

We extend Isabelle/ITrees to support real valued vectors and matrices (Section~\ref{sec:vcg}) and use it to prove the total correctness of four numerical algorithms: the bisection method (Section~\ref{sec:bisection}), the fixed-point method (FPM - Section~\ref{sec:formalise-fpm}), the perceptron (Section~\ref{sec:perceptron}), and the gradient descent algorithm (Section~\ref{sec:grad-descent}). To achieve this, we substantially extend Isabelle's Analysis libraries with generalisations of higher-order derivatives and Taylor's formula with a Peano remainder. Our experience is generally positive, requiring the following steps for each numerical method: (1) model the algorithm in Isabelle/ITrees, (2) experiment with the algorithm by executing it, (3) write its correctness specifications, (4) pose a suitable set of invariants and variants, (5) use the VCG automation to produce proof obligations, (6) if required, prove theorems from Analysis to aid in discharging the proof goals and (7) repeat from step 4 until the postcondition is proved. We conclude our paper by discussing related (Section~\ref{sec:related}) and future work (Section~\ref{sec:concl}). To our knowledge, this is the first application of an imperative verification framework to a diverse set of numerical algorithms.



While this paper focuses on a version of these algorithms using arbitrary precision, we have also explored a generative AI-based autoformalisation approach~\cite{bryantHKU2026} to adapt our manual verifications to floating-point arithmetic. 
We believe that the formalisation of these algorithms (and others omitted due to space constraints) via Isabelle/ITrees and the Isabelle/HOL libraries cements these tools as a foundation for verifying numerical algorithms.  Our formalisations are available via GitHub and links to them appear throughout. \hfill\isalink{https://github.com/brydustin/Verification_Of_One_Dim_Optimization/tree/036eea319e62b5d65d809fcf9ca2defd9f312f80}

\section{Isabelle/ITrees}
\label{sec:vcg}

We describe relevant features of the Isabelle/ITrees library~\cite{Foster2024ITrees} and showcase the extensions that we implemented to enable its support for numerical methods. 

Isabelle/ITrees is part of the Isabelle/UTP semantics and verification ecosystem~\cite{Foster2020-IsabelleUTP}. The library provides an abstract imperative programming notation, mechanisms for total-correctness specifications, (in)variant annotations, and a verification-condition generator. The programs are assigned a formal semantics in terms of Interaction Trees (ITrees), which unify denotational and operational semantics. In particular, ITrees, as coinductively defined data structures, are executable in nature, and the library leverages Isabelle's code generator for this. We extend the existing tool to support real number variables by using the \textsf{HOL-Analysis} library and providing implementations of vector and matrix operations to the code generator, enabling the execution of numerical algorithms.

Consider a program for scalar-vector multiplication in Isabelle/ITrees: \hfill\isalink{https://github.com/isabelle-utp/Verified_Numerical_Methods/blob/b4b3aec6a67ab28e4c6a638c02e47702ab9d1834/Vector_Scale.thy\#L48}

\begin{lstlisting}[style=isabellestyle, language=Isabelle]
alphabet ('n :: finite) st = i::nat  vc::"real vec['n]"

program vec_scale "(c :: real, X ::real vec['n])" =
"vc := X; i := 0; 
 while i < CARD('n) do vc[i] := c * X(i); i := i + 1 od"
\end{lstlisting}

\noindent Here, the \lstinline{alphabet} command declares the program's state space, consisting of two variables: the natural number \texttt{i}, and the (real) vector \texttt{vc}. The state type \texttt{st} is parametric over type \texttt{'n}, which is the vector's dimension of finite cardinality.

We define program \texttt{vec\_scale} using the \lstinline{program} keyword.
The program takes as input the (real) scalar \texttt{c}, and a vector \texttt{X} of dimension \texttt{'n}. Having initialised \texttt{vc} and \texttt{i} to \texttt{X} and $0$ respectively, the program iterates through each element of \texttt{vc}, performing a point-wise multiplication. The notation \texttt{CARD('n)} gives the dimension of the vector as a natural number. The operation \texttt{X(i)} retrieves the \texttt{i}th element of \texttt{X}. We can run the program using the \lstinline{execute} keyword:

\begin{lstlisting}[style=isabellestyle, language=Isabelle]
execute "vec_scale(3, Vector[1,2,3,4])"
\end{lstlisting}

\noindent The \texttt{Vector} notation builds a vector of the inferred dimension (4). 
The \lstinline{execute} command compiles the \texttt{vec\_scale} program to an ITree, generates SML code, and then executes the code. Isabelle reports that the program terminates with state \texttt{i = 4, vc = Vector[3, 6, 9, 12]}. Execution is feasible only when the real number inputs are rational, and so supported by multi-precision arithmetic.

We can verify such a numerical algorithm using Hoare logic. The Hoare triple below states a total correctness claim that the program terminates and, in the final state, \texttt{vc} is the result of a scalar multiplication, denoted in Isabelle via $*_R$.

\begin{lstlisting}[style=isabellestyle, language=Isabelle]
lemma "H[True] vec_scale(c, X) [vc = c ~$*_R$~ X]"
\end{lstlisting}

\noindent  Isabelle/ITrees includes the \texttt{vcg} proof method, which applies Hoare logic laws to produce verification conditions. This often requires annotating the program with loop (in)variants~\cite{Foster2020-IsabelleUTP,Foster2024ITrees,HuertaFGSLH2024}.
Below, we add to \texttt{vec\_scale} both,
\hfill{\isalink{https://github.com/isabelle-utp/Verified_Numerical_Methods/blob/b4b3aec6a67ab28e4c6a638c02e47702ab9d1834/Vector_Scale.thy\#L56}}

\begin{lstlisting}[style=isabellestyle, language=Isabelle]
while i < CARD('n) 
  invariant i ~≤~ CARD('n) ~∧~ (~∀~k<i. vc k = c * X k)
  variant CARD('n) - i do ... od
\end{lstlisting}

\noindent where the dots (\texttt{...}) abbreviate the above body of the while-loop. The \texttt{vcg} method yields the extensionality principle for vectors as a proof obligation.
Namely, if each element of the vector \texttt{vc} is \texttt{c} times the corresponding element of \texttt{X}, then \texttt{vc} is the scalar multiple \lstinline{c $*_R$ X}. This VC can be discharged automatically by application of Isabelle's \textsf{Sledgehammer} tool.

\section{Bisection Method}
\label{sec:bisection}

We begin assessing Isabelle/ITrees' suitability for verifying numerical algorithms with the bisection method—a classic, derivative-free root-finding algorithm. We choose it because it is an accessible entry point to the field. We describe its implementation, justify the correctness specification and loop invariants, and discuss the challenges in the corresponding proof.


Recall from Bolzano's Theorem that a real-valued function $f$, continuous on a closed interval $[a,b]\subseteq\mathbb{R}$, and such that $f(a)$ and $f(b)$ have opposite signs, has a root inside the open interval $(a,b)$. The bisection method iterates this aforementioned fact (see Listing~\ref{list:bisect}): at each step $n$, it defines the current bracket as an interval, $[l_n,u_n]$, that halves the previous bracket (where $[l_0,u_0]=[a,b]$). The bisection method evaluates $f$ at the bracket's midpoint $m_n=(l_n+u_n)/2$, and retains the bracket's half on which the sign change persists. It repeats this process until the bracket's length falls below a given threshold. This halving process explains the term \emph{bisection}. The algorithm is formalised below:\hfill\isalink{https://github.com/brydustin/Verification_Of_One_Dim_Optimization/blob/036eea319e62b5d65d809fcf9ca2defd9f312f80/Numerical_Methods/Bisection.thy\#L116}


\begin{lstlisting}[style=isabellestyle, language=Isabelle, caption= Bisection method in Isabelle, label={list:bisect}] 
program bisection "(f::real~⇒~real, a::real, b::real, tol::real)" 
  over state = "iter:= 0; fa:= f(a); fb:= f(b); l:= a; u:= b; 
  xmid:= l; ymid:= f(xmid);         
    while u-l>tol do 
      iter:= iter+1; xmid:= (l+u)/2; ymid:= f(xmid); 
      if fa*ymid>0 then l:= xmid; fa:= ymid 
      else u:= xmid; fb:= ymid fi od"
\end{lstlisting}



We initialise the bracket $[\texttt{l},\, \texttt{u}]$ to $[a,b]$, while \texttt{fa} and \texttt{fb} denote the value of the input function \texttt{f} at the bracket's left and right limits. The \lstinline{while} loop checks that the bracket's size is strictly larger than the input tolerance (\texttt{u-l>tol}), while the loop's body computes the current bracket's midpoint \texttt{xmid:=(l+u)/2} and the value of \texttt{f} at that midpoint, \texttt{ymid}. The chosen half of the bracket depends on the sign agreement of \texttt{f} at the left endpoint and midpoint (\texttt{fa*ymid > 0}).



Having discussed the intuition for the algorithm's behaviour, we state the bisection method's total correctness below and explain our assumptions.\hfill\isalink{https://github.com/brydustin/Verification_Of_One_Dim_Optimization/blob/036eea319e62b5d65d809fcf9ca2defd9f312f80/Numerical_Methods/Bisection.thy\#L199}

\begin{lstlisting}[style=isabellestyle, language=Isabelle, caption= Total correctness of the bisection method, label={list:correct-bisect}] 
theorem bisection_error_bound:
  assumes "0 < tol" "tol < b-a" 
    "f(a)*f(b) < 0" "continuous_on {a..b} f" 
  shows "H[True] bisection(f,a,b,tol) [
    ~∃~c. f(c) = 0 ~∧~ c ~∈~ {a<..<b} ~∧~ u-l ~≤~ tol
    ~∧~ ~¦~c - xmid~¦~ ~≤~ (b-a)/2^iter 
    ~∧~ ~¦~c - xmid~¦~ ~≤~ tol ~∧~ iter = ~⌈~log 2 (b-a)/tol~⌉~]"
\end{lstlisting}

\noindent The theorem in Listing~\ref{list:correct-bisect} asserts that \texttt{bisection} terminates, returning \texttt{xmid} as an approximation to a root of \texttt{f}.  That is, $\texttt{xmid}$ is at most a distance of $\texttt{tol}$ to a root $\texttt{c}$ after precisely $\left\lceil \log_2\left(\frac{\texttt{b}-\texttt{a}}{\texttt{tol}}\right)\right\rceil$ iterations, provided \texttt{f} is continuous on $[a,b]$.  We assume continuity on $[a,b]$ and $\texttt{f(a)*f(b)<0}$ so that we may apply Bolzano's theorem.
The assumptions \texttt{tol<b-a} and \texttt{0<tol} guarantee a nonempty initial domain and the possibility of termination.
Notably, our implementation of the method does not terminate if $\texttt{f(xmid)=0}$.

Finding the right variant was crucial (and challenging) to prove the termination of the bisection method.
The program's loop (in)variants are below.\hfill\isalink{https://github.com/brydustin/Verification_Of_One_Dim_Optimization/blob/036eea319e62b5d65d809fcf9ca2defd9f312f80/Numerical_Methods/Bisection.thy\#L119}

\begin{lstlisting}[style=isabellestyle, language=Isabelle]
  invariant fa = f(l) ~∧~ fb = f(u) ~∧~ (l = xmid ~∨~ u = xmid) 
  ~∧~ a ~≤~ l ~∧~ u ~≤~ b ~∧~ l < u ~∧~ fa*fb ~≤~ 0 
  ~∧~ u-l = (b-a)/2^iter ~∧~ 2*(u-l) > tol
  variant nat(~⌈~log 2 (b-a)/tol~⌉~) - iter
\end{lstlisting}

\noindent The conjuncts \texttt{fa=f(l)} and \texttt{fb=f(u)} state that the cached values \texttt{fa} and \texttt{fb} agree with those of the function at the bracket's endpoints
\texttt{l} and \texttt{u}.
The disjunction $\texttt{l = xmid} \,\vee\, \texttt{u = xmid}$ declares that
each iteration replaces one of the interval's endpoints by the midpoint \texttt{xmid}, corresponding to either the \lstinline{then} or \lstinline{else} branch in
Listing~\ref{list:bisect}.
The bounds $\texttt{a}\le \texttt{l}$, $\texttt{l}< \texttt{u}$, and $\texttt{u}\le \texttt{b}$ state that the bracket remains a nonempty
subinterval of \([a,b]\).
The sign condition $\texttt{fa*fb}\le\texttt{0}$ preserves applicability of Bolzano's theorem; again we note that since possibly \texttt{f(xmid)=0} it could be the case that  \texttt{fa*fb}=\texttt{0}. Bolzano is only applicable if $\texttt{fa*fb}<\texttt{0}$, but fortunately \texttt{vcg} handles these cases separately. Finally, the invariant $\texttt{u-l=(b-a)/}2^{\texttt{iter}}$ captures the bracket's halving at each iteration. 

From the above invariants, one can prove that the program's iterations \texttt{iter} are bounded below by $\left\lceil \log_2\!\left(\frac{\texttt{b}-\texttt{a}}{\texttt{tol}}\right)\right\rceil$. To derive the exact value for \texttt{iter} (to derive the variant), we observe that the while loop's condition is always respected in the predecessor step and the bracket's size is always halved: $\texttt{tol}<u_{i-1}-l_{i-1}=2\cdot(u_i-l_i)$, which remains true when \texttt{iter=0} by the assumptions \texttt{0<tol<b-a}. From this observation, the violation of the loop's condition at the last iteration and the second-to-last invariant, the number of iterations is $\left\lceil \log_2\!\left(\frac{\texttt{b}-\texttt{a}}{\texttt{tol}}\right)\right\rceil$.

With the aforementioned loop (in)variants, Isabelle/ITrees' \texttt{vcg} method generates 15 goals in total, 13 of which can be proven directly with Isabelle's \textsf{Sledgehammer}. The remaining two proof obligations are symmetrical and related to the last invariant. With algebraic manipulations, one can discharge them.

Overall, the formalisation and verification of the bisection method required 195 lines of Isabelle code under the counting convention of Section~\ref{sec:eval}. Notably, Isabelle/ITrees enables the straightforward formalisation of the algorithm, its correctness specification, and loop (in)variants. Furthermore, the framework's automation reduces the bisection method's proof of correctness to a single goal, up to symmetry, proven by application of Bolzano's Theorem and some algebra. Thus, the verification process for this program is highly automated.

\section{Fixed-Point Method}
\label{sec:formalise-fpm}

We now focus on a numerical method with stronger mathematical prerequisites, the fixed-point method (FPM). As before, we discuss the algorithm's formalisation, its correctness specification, and the challenges and intuitions in proving it. In particular, our analysis of the conditions under which the FPM is applicable requires expanding the libraries of formalised mathematics in Isabelle with an implementation of higher-order differentiability and Taylor's theorem.

As the name hints, the FPM approximates a \emph{fixed point} of its input function $f$. These are points $r$ such that $f(r)=r$. It is also a root-finding method since $f(r) = r$ is equivalent to $f(r)-r=0$. The FPM starts from an initial guess $x_0$ and repeatedly applies $f$, generating a sequence $x_{n+1}=f(x_n)$  converging to $r$. Informally, its convergence relies on assuming that the input function $f$ reduces the distance between points in a neighbourhood $S$ of $x_0$ after each application. Formally, it relies on $f$ satisfying the conditions of the Banach fixed-point theorem, that is, $S$ should be a contractive neighbourhood of the initial point $x_0$ under $f$ (namely, $\lvert f(x) - f(y) \rvert \le c\cdot\,\lvert x - y \rvert$ for all $x,y\in S$, where $c<1$). As before, the program's formalisation is straightforward in the framework:\hfill\isalink{https://github.com/brydustin/Verification_Of_One_Dim_Optimization/blob/036eea319e62b5d65d809fcf9ca2defd9f312f80/Numerical_Methods/Fixed_Point_Method.thy\#L447}

\begin{lstlisting}[style=isabellestyle, language=Isabelle, caption= Fixed-point method in Isabelle, label={list:fpi}] 
program fpm "(f::real~⇒~real , x0::real , tol::real)" over st
  ="x := x0; x_new := f x; iter := 0;
  while (~¦~x_new - x~¦~ ~≥~ tol) do 
    x := x_new; x_new := f x; iter := iter + 1 od"
\end{lstlisting}

\noindent In Listing~\ref{list:fpi}, the variables \texttt{x} :: $\mathbb{R}$ and \texttt{x\_new} :: $\mathbb{R}$ correspond to $x_n$ and $x_{n+1}$ respectively, where $n$ is the number of iterations \texttt{iter} :: $\mathbb{N}$.  The distance test $\lvert \texttt{x\_new} - \texttt{x} \rvert \ge \texttt{tol}$ controls the loop's termination. Irrespective of this test, the loop's body executes the assignments \texttt{x := x\_new}, \texttt{x\_new:= f x}, and \texttt{iter:= iter + 1}. The loop exits when the latest distance falls below the tolerance. A first, total correctness theorem for the fixed point method appears below.\hfill\isalink{https://github.com/brydustin/Verification_Of_One_Dim_Optimization/blob/036eea319e62b5d65d809fcf9ca2defd9f312f80/Numerical_Methods/Fixed_Point_Method.thy\#L686}

\begin{lstlisting}[
  style=isabellestyle,
  language=Isabelle,
  caption={Total correctness of the FPM in a contractive neighbourhood},
  label={list:correct-fpi}
]
lemma fpm_iter_bnd_local_leq:
  assumes "0 ~≤~ c" "c < 1" "0 < tol" "0 < ~δ~" "f r = r"
    and "~¦~x0 - r~¦~ ~≤~ ~δ~"
    and "~∀~s t. ~¦~s - r~¦~ ~≤~ ~δ~ ~∧~ ~¦~t - r~¦~ ~≤~ ~δ~ 
      ~⟶~ ~¦~f s - f t~¦~ ~≤~ c*~¦~s - t~¦~"
  shows "H[True] fpm(f, x0, tol)[
  ~¦~x-r~¦~~≤~c ^ iter * ~¦~x0-r~¦~ ~∧~ ~¦~x_new-x~¦~<tol 
  ~∧~ iter ~≤~ Suc (nat~⌈~ln (max 1 (~¦~f x0 - x0~¦~/tol))/ln (1/c)~⌉~)]" 
\end{lstlisting}

The first two assumptions in Listing~\ref{list:correct-fpi} specify the contraction constant, while the last assumption characterises contractivity. We choose an interval of radius $\delta > 0$ as the contractive neighbourhood of the initial guess and assume that a fixed point of $f$ exists in this interval. We also proved a version of the aforementioned lemma that replaces $\le$ in the assumption $|x_0-r|\le \delta$ with $<$, and similarly replacing it in the antecedent of the final assumption.  Below, we write our loop invariants and variant for the FPM:
\begin{lstlisting}[style=isabellestyle, language=Isabelle] 
invariant x = (f^^iter) x0 ~∧~ x_new = f x 
  ~∧~ ~¦~x_new-x~¦~ ~≤~ (c^iter)*~¦~f x0-x0~¦~
  ~∧~ iter ~≤~ Suc nat~⌈~ln(max 1 (~¦~f x0 - x0~¦~/tol))/ln (1/c)~⌉~)   
 variant  Suc(nat~⌈~ln(max 1 (~¦~f x0 - x0~¦~/tol))/ln (1/c)~⌉~) - iter
\end{lstlisting}

\noindent The first invariant \texttt{x = f$^\texttt{iter}\ $x0} specifies that, at iteration \texttt{iter}, the variable \texttt{x} holds the \texttt{iter}-th application of $f$ to the initial guess, and \texttt{x\_new = f x} relates each iteration to the next one. Also, contractivity implies that the step sizes decay geometrically, thus $\lvert x_{k+1} - x_k \rvert \le c^k \,\lvert f(x_0) - x_0 \rvert$. To prove termination without an explicit \texttt{max\_iter}, we derive a closed-form iteration bound from the loop guard and apply induction on $k$. Notice that $c$ is missing from the program. We can reason about it from the (in)variants with a renaming aid where we define \texttt{fpm\_aux(f, x0, tol,c)} (which receives the ghost variable $c$), and prove it equivalent to the original: \texttt{fpm\_aux(f, x0, tol,c)} $=$ \texttt{fpm(f, x0, tol)}.

After applying \texttt{vcg}, four proof obligations emerge: one discharged by \textsf{Sledgehammer}, one requiring an induction, and the remaining needing straightforward algebraic manipulations with intermediate steps discharged by \textsf{Sledgehammer}.

As stated above, the total correctness theorem requires its users to identify the contraction constant $c$ for $f$ on a neighbourhood around the fixed point. To alleviate these burdens, we use the rest of the Section to formalise two results that establish conditions under which a contractive neighbourhood exists. For this, we closely follow Solomon's presentation~\cite{Solomon2015NumericalAlgorithms}. 

Firstly, we consider a continuously differentiable function $f\in C^1(U)$ such that $\lvert f'(r)\rvert<1$.
We can find an $\varepsilon>0$, such that $1-\lvert f'(r)\rvert-\varepsilon>0$. Then by continuity of $f'$, 
there is also a $\delta>0$ such that, if $\lvert x -r\rvert<\delta$, then $1-\varepsilon-\lvert f'(r)\rvert > \lvert f'(x)-f'(r)\rvert$. 
By the triangle inequality, $\lvert f'(x)\rvert<1-\varepsilon$ holds in a neighbourhood of radius $\delta$ around the fixed point $r$. Then, by the mean value theorem $\lvert f(x)-f(r)\rvert = \lvert f'(\theta)\rvert\lvert x-r\rvert$ for some $\theta$ if $x\in[r-\delta,r+\delta]$. Hence, $\lvert f(x)-f(r)\rvert < (1-\varepsilon)\lvert x-r\rvert$. Thus, users only need to provide a $C^1$ function and an initial guess $x_0$ close to the fixed point. The formal statement appears below, and as before, we may replace the $\le$ in the goal's antecedent with $<$. \hfill\isalink{https://github.com/brydustin/Verification_Of_One_Dim_Optimization/blob/036eea319e62b5d65d809fcf9ca2defd9f312f80/Numerical_Methods/Fixed_Point_Method.thy\#L817}

\begin{lstlisting}[style=isabellestyle, language=Isabelle, caption= Convergence when $f\in C^1(U)$ and $|f'(r)|<1$,
  label={list:correct-c1-fpi}] 
lemma fpm_known_iter_bnd_C1_leq:
assumes "f r = r" "r ~∈~ U" "C_k_on 1 f U"
  and "~¦~deriv f r~¦~ < 1" "0 < tol" 
shows  "~∃~ ~δ~ > 0. ~∃~ ~ε~ > 0. (~∀~(x0 :: real). ~¦~r - x0~¦~ ~≤~ ~δ~ 
  ~⟶~ H[True] fpm(f, x0, tol) [~¦~x - r~¦~ ~≤~ (1 - ~ε~)^iter*~¦~x0 - r~¦~ 
    ~∧~ ~¦~x_new - x~¦~ < tol ~∧~ 
    iter ~≤~ Suc(nat~⌈~ln(max 1 (~¦~f x0 - x0~¦~/tol))/ ln(1/(1-~ε~))~⌉~)])"  
\end{lstlisting}

The second case that we consider requires the first derivative of $f$ at the fixed point to be $0$ ($f'(r)=0$), and the second derivative $f''(r)$ to exist. 
Since $f$ is twice differentiable, by Taylor's theorem in its Peano remainder form, there is a function $h$, tending to $0$ at $x_n$, such that
$$f(x_{n})= f(r) + f'(r)\cdot(x_n-r)+ \frac{f''(r)}{2}\cdot (x_n-r)^2+\frac{h(x_n)}{2}\cdot(x_n-r)^2.$$
We can replace $h(x_n)$ by an appropriate $\varepsilon>0$ and use the triangle inequality to derive the following bound on an iterate of the FPM:
$$\lvert x_{n+1}-r\rvert = \lvert f(x_n)-f(r)\rvert\leq \frac{1}{2}\lvert f''(r)+\varepsilon\rvert\cdot (x_n-r)^2.$$

\noindent That is, the distance between the next iteration and the desired fixed point reduces quadratically. This is often highlighted to indicate that under the above-stated conditions, the FPM converges faster (quadratically). We have also formalised this result and proven it using features of the Isabelle/ITrees framework.\hfill\isalink{https://github.com/brydustin/Verification_Of_One_Dim_Optimization/blob/036eea319e62b5d65d809fcf9ca2defd9f312f80/Numerical_Methods/Fixed_Point_Method.thy\#L1124}

\begin{lstlisting}[style=isabellestyle,language=Isabelle,
caption={Quadratic conv. when $f \in C^1(U)$, $f$ twice diff. at $r$, and $f'(r)=0$},
  label={list:correct-taylor-fpi}]
lemma fpm_iter_quadratic_convergence:
  assumes"f(r) = r" "r ~∈~ U" "C_k_on 1 f U" 
    and "deriv f r = 0" "0 < tol" "f twice_differentiable_at r"
  shows "~∃~(~δ~ :: real)>0. ~∃~(~ε~ :: real)>0. (~∀~x0. ~¦~r - x0~¦~ < ~δ~ 
    ~⟶~ H[True] fpm(f, x0, tol) [~¦~x_new - x~¦~ < tol ~∧~ ~¦~x-r~¦~ 
    ~≤~ (((~¦~deriv(deriv f) r~¦~ + ~ε~)/2)^(2^iter-1))*~¦~x0-r~¦~^(2^iter) 
~∧~ iter ~≤~ Suc(nat~⌈~ln(max 1 (~¦~f x0 - x0~¦~/tol)) / ln(1/(1-~ε~))~⌉~)])"  
\end{lstlisting}

Unlike the bisection method, the formalisation and verification of these results are not straightforward. Concretely, the goals that emerged from the \texttt{vcg} method on the specification in Listing~\ref{list:correct-fpi} are not automatically discharged with \textsf{Sledgehammer}. Yet, their proofs are still relatively simple, requiring simple inductions with intermediate steps that \textsf{Sledgehammer} can prove. However, our formalisation and verification of the specifications in Listings ~\ref{list:correct-fpi}, ~\ref{list:correct-c1-fpi}, and ~\ref{list:correct-taylor-fpi} require several, long $\varepsilon-\delta$ arguments, making the fixed-point development span 1111 lines of Isabelle code under the counting convention of Section~\ref{sec:eval}. This number does not account for the extensions to the libraries of formalised mathematics in Isabelle that we needed to state our correctness specifications. Namely, to assert the conditions related to the higher-order differentiability of the input function $f$ and the possibility of approximating $f$ via a Taylor expansion, the existing libraries were insufficient. In particular, to our knowledge, we contribute (1) the most general definition of higher-order differentiability in any Isabelle library (all other concepts in the AFP either imply ours or are instances of ours) and (2) the only formalisation of Taylor's theorem in its Peano form. These contributions, along with further concepts from vector calculus, span 4754 lines of Isabelle code under the same convention.



\section{Perceptron}
\label{sec:perceptron}

The perceptron is a supervised learning algorithm for binary classification, which seeks parameters $\texttt{theta}\in\mathbb{R}^n$ and $\texttt{bias}\in\mathbb{R}$ such that
$\textrm{sgn}(x_i\cdot\texttt{theta} + \texttt{bias}) = y_i$ for each labelled example $(x_i,y_i)$ in a dataset $D$, where $y_i\in\{-1,+1\}$ is the label of $x_i\in\mathbb{R}^n$. The algorithm is a semi-decision procedure for linear separability: it terminates after finitely many iterations with a pair $(\texttt{theta}, \texttt{bias})$ that correctly classifies all examples if and only if the data is linearly separable.  Geometrically, $H\, {(\theta,b)} = \{x| x\cdot \theta + b = 0\}$ defines a hyperplane with $\theta$ as a normal vector and the objective is to find $(\texttt{theta},\texttt{bias})$ such that $H\, {(\texttt{theta},\texttt{bias})}$ strictly divides the two classes with a positive margin $\gamma > 0$. Below, we present the algorithm in two stages: \texttt{pass\_through}, which ``learns'' from the data, and \texttt{perceptron}, which repeatedly iterates \texttt{pass\_through} until no misclassifications are detected. 

\begin{lstlisting}[style=isabellestyle,language=Isabelle,
  caption={One pass (epoch) of the perceptron over the training data.},
  label={lst:perceptron-pass}]
program pass_through "D :: ((real vec['n]) ~×~ real) list"
  over "('n :: finite) state"
  ="changed:= False; i:= 0;
  while i < length D do x_i:= fst (D!i); y_i:= snd (D!i);
    if y_i*(x_i~∙~theta+bias)~≤~0 then theta:= theta+y_i*$_R$x_i; bias:= bias+y_i; 
    updates:= updates+1; changed:= True fi; i:= i+1 od"
\end{lstlisting}

\begin{lstlisting}[style=isabellestyle,language=Isabelle,
  caption={The classic perceptron as repeated passes until an epoch makes no updates.},
  label={lst:perceptron-main}]
program perceptron "D :: ((real vec['n]) ~×~ real) list"
over "('n :: finite) state"
= "theta:= 0; bias:= 0; changed:= True; updates:= 0;
 while changed do pass_through D od"
\end{lstlisting}

The variable \texttt{updates::nat} counts the number of changes to the ordered pair $(\texttt{theta},\texttt{bias})$, while \texttt{changed::bool} records if \texttt{pass\_through} updated the pair at least once. In such a case, the \texttt{perceptron} will call \texttt{pass\_through} again. The updates are assignments $\texttt{theta}:=\texttt{theta}+y_i*x_i$ and $\texttt{bias}:= \texttt{bias}+y_i$ and occur only when the sign of $y_i$ and $x_i\cdot \texttt{theta}+\texttt{bias}$ do not match. That is, when $x_i$ is on the wrong side of $H\, {(\texttt{theta},\texttt{bias})}$.

The perceptron convergence theorem (PCT) says that if the data are linearly separable with margin $\gamma>0$, and all the datapoints $(x_i, y_i)$ are inside the circle of radius $R>0$ ($\|x_i\|^2 + 1 \le R^2$), then the perceptron makes at most $(R/\gamma)^2$ updates before it finds parameters $(\texttt{theta},\texttt{bias})$ that correctly classify all points in $D$.
We formalise the PCT as:

\begin{lstlisting}[style=isabellestyle,language=Isabelle,
  caption={The perceptron convergence theorem},
  label={lst:perceptron-convergence}]
theorem perceptron_convergence_theorem:
  assumes "~γ~ > 0" "~∥~~θ~~∥~$^2$ + $b^2$ = 1" 
    and "~⋀~t. t < length D ~⟹~
    ~∥~fst(D!t)~∥~$^2$ + 1 ~≤~ R$^2$ ~∧~ snd(D!t)~∈~{-1, 1} ~∧~ snd(D!t)*(fst(D!t)~∙~~θ~+b) ~≥~ ~γ~"
  shows "H[True] perceptron (D) [real updates ~≤~ (R/~γ~)$^2$ 
    ~∧~ (~∀~i < length D. sgn(fst(D!i) ~∙~ theta + bias) = snd(D!i))]"
\end{lstlisting}

The assumption about separability is an existential assertion of a hyperplane $H(\theta, b)$. To address this, we define the program \texttt{perceptron\_aux\ $(D,\theta, b,\gamma,R)$} (with its respective pass-through) that receives as inputs the ghost variables $R,\gamma, \theta$, and $b$, but that is otherwise indistinguishable from the original perceptron program. In our formalisation, we perform the invariant reasoning on this auxiliary program and show that introducing these new variables does not change the semantics: \texttt{perceptron\_aux\ $(D,\theta, b,\gamma,R) = $ perceptron\ $(D)$}. The loop variants and invariants of \texttt{pass\_through\_aux} and \texttt{perceptron\_aux} are shown below.

\begin{lstlisting}[style=isabellestyle,language=Isabelle,
  caption={Loop (in)variants of \texttt{pass\_through\_aux}},
  label={lst:perceptron-pass-aux-inv}]
invariant updates*~γ~ ~≤~ (~θ~~∙~theta)+(b*bias) ~∧~ ~∥~theta~∥~$^2$ + bias$^2$ ~≤~ updates*R$^2$
 ~∧~ (changed ~⟶~ old[updates]<updates) ~∧~ (~¬~changed ~⟶~ updates = old[updates])
 ~∧~ (~¬~ changed ~⟶~ (~∀~j < i. (snd(D!j))*((fst(D!j)) ~∙~ theta + bias) > 0)) 
variant length D-i
\end{lstlisting}

\begin{lstlisting}[style=isabellestyle,language=Isabelle,
  caption={Loop (in)variants of \texttt{perceptron\_aux}},
  label={lst:perceptron-aux-inv}]
invariant updates*~γ~ ~≤~ (~θ~~∙~theta)+(b*bias) ~∧~ ~∥~theta~∥~$^2$ + bias$^2$ ~≤~ updates*R$^2$
 ~∧~ (~¬~ changed ~⟶~(~∀~j < length D. (snd(D!j))*((fst (D!j)) ~∙~ theta+bias) > 0))
 ~∧~ updates ~≤~ nat ~⌈~(R/~γ~)$^2$~⌉~
variant (nat ~⌈~(R/~γ~)$^2$~⌉~ - updates + (if changed then 1 else 0))
\end{lstlisting}

The first two invariants characterise the perceptron's standard proof of convergence~\cite{MurphyGS2017}. Namely, one must inductively show that the angle between the hyperplanes $H(\texttt{theta},\texttt{bias})$ and $H(\theta,b)$ is bounded from below by the updates times the margin, and that the norm of the updated normal vector never grows larger than the radius $R$. From these, one deduces the maximum number of updates $(R/\gamma)^2$ required for convergence of the perceptron (which is itself an invariant). Additionally, the Isabelle/ITrees framework includes an operator \texttt{old} enabling us to describe the state of a variable before its most recent assignment. Thanks to this helpful feature, the remaining invariants denote the above-described behaviour of \texttt{updates} and $H(\texttt{theta},\texttt{bias})$ relative to the \texttt{changed} variable. The separation of the programs enables us to carry only the necessary invariants from the inner loop to the outer one. For the variants, the first one says that each iteration in a pass through the data decreases the number of data points to visit. The second one coincides with the main invariant, except for the fact that one has to consider that the program requires an extra pass through the data to certify that everything has been classified correctly. Thus, in the final iteration, there are no errors and the program terminates, but since \texttt{updates} does not change, the variants should still be decreased, which is expressed with the term \texttt{(\lstinline{if} changed \lstinline{then} 1 \lstinline{else} 0)}.

\section{Gradient Descent}
\label{sec:grad-descent}

The gradient descent algorithm stands out among multidimensional optimisation methods for its simplicity and prominence in modern machine-learning practice. The method is so-named as it updates a vector approximation \texttt{x} ::\texttt{~ℝ~vec['i]} to a minimiser \texttt{x\_min} (a point $x^*$ such that $f(x^*)\leq f(x)$ for all $x$), by moving in the direction opposite to the steepest ascent. That is, it moves opposite to the gradient \texttt{grad(x)} of \texttt{f} at \texttt{x}, a distance of $\alpha$\texttt{(x)>0} at each step. The method exits its while-loop when the first-order optimality condition \texttt{grad(x)} $\bullet$ \texttt{grad(x)} $\le \varepsilon^2$ is satisfied. We showcase the formalisation of the method below.\hfill\isalink{https://github.com/brydustin/Verification_Of_One_Dim_Optimization/blob/036eea319e62b5d65d809fcf9ca2defd9f312f80/Numerical_Methods/Gradient_Descent.thy\#L616}

\begin{lstlisting}[style=isabellestyle,language=Isabelle,
  caption={Gradient descent algorithm},
  label={lst:gradient_descent}]
program gradient_descent "(grad :: ~ℝ~ vec['i] ~⇒~ ~ℝ~ vec['i], x0 :: ~ℝ~ vec['i], ~α~ :: ~ℝ~ vec['i] ~⇒~ ~ℝ~, ~ε~ :: ~ℝ~)" 
  over "'i::finite st" = "x:= x0; iter:= 0; 
  while (grad(x) ~∙~ grad(x) > ~ε~$^2$) do x:= x-(~α~ x)*$_R$ grad(x); iter:= iter+1 od"
\end{lstlisting}

\noindent The choice of step length $\alpha$\texttt{(x)} in general depends on \texttt{x}, and on the optimisation problem. However, there is no general-purpose method for specifying its value, and in general, the method may not converge~\cite{NocedalW2006}.

We verified \emph{three} versions of gradient descent, all sharing the program of Listing~\ref{lst:gradient_descent} and differing only in their assumptions on $f$, their convergence rate, and the resulting iteration bound. Ordered from the weakest to the strongest hypotheses, these are: (i) a general, possibly nonconvex, $L$-smooth objective, for which the method reaches an $\varepsilon$-stationary point in $O\!\left(L\,(f(x_0)-f(x^*))/\varepsilon^2\right)$ steps; (ii) an $L$-smooth objective satisfying the Polyak-Łojasiewicz (PL) inequality with the constant step $1/L$, which converges \emph{linearly}; and (iii) a strongly convex quadratic $f(x)=\tfrac12\,x^\top Q x - b^\top x$ minimised by exact line search, which converges at the Kantorovich rate $\bigl(\tfrac{A-a}{A+a}\bigr)^2$, where $0<a\le A$ bound the eigenvalues of $Q$. As in the fixed-point method, each result is discharged for the imperative program by the \texttt{vcg} method, with the problem data ($L$, $\mu$, the minimiser, and the eigenvalue bounds) supplied to the proof through an auxiliary program proven equal to \texttt{gradient\_descent}.

The first regime assumes no convexity and only an $L$-Lipschitz gradient. The descent lemma then guarantees a fixed energy decrease per step while $\|\nabla f(x)\|>\varepsilon$, and telescoping this decrease bounds the number of iterations needed to reach an $\varepsilon$-stationary point.\hfill\isalink{https://github.com/brydustin/Verification_Of_One_Dim_Optimization/blob/036eea319e62b5d65d809fcf9ca2defd9f312f80/Numerical_Methods/Gradient_Descent.thy\#L785}

\begin{lstlisting}[style=isabellestyle,language=Isabelle,
  caption={Convergence to an $\varepsilon$-stationary point under $L$-smoothness alone},
  label={lst:gd-general-conv}]
theorem gradient_descent_general_convergence:
  assumes "0 < ~ε~" "0 < L" "~⋀~x. ~α~ x = 1 / L"
   and "~⋀~x. f x_min ~≤~ f x" "~⋀~x. (f has_derivative (~λ~h. grad x ~∙~ h)) (at x)"
   and grad_Lipschitz: "~⋀~x y. ~∥~grad x - grad y~∥~ ~≤~ L*~∥~x - y~∥~"
  shows "H[True] gradient_descent (grad, x0, ~α~, ~ε~)
  [~∥~grad x~∥~ ~≤~ ~ε~ ~∧~ iter ~≤~ nat ~⌈~2*L*(f x0 - f x_min)/~ε~$^2$~⌉~]"
\end{lstlisting}

\noindent The second regime keeps the constant step $\alpha$\texttt{(x)}$=1/L$ and the $L$-Lipschitz gradient, and additionally assumes that $f$ satisfies the Polyak-Łojasiewicz  property:
$$\frac{1}{2} \| \nabla f(x)\|^2 \ge \mu\cdot (f(x)-f(x^*)),\quad\text{ for all }x.\qquad (\textrm{PL})$$

\noindent Intuitively this indicates that ``if $f(x)$ is far from $f(x^*)$, then $x$ is not at a flat plateau because the gradient squared $\|\nabla f(x)\|^2$ must be greater''. Moreover, if $\nabla f(x) = 0$ then $f(x)=f(x^*)$. Under these conditions, starting from \texttt{x0}, the $f$ converges linearly to \texttt{f(x\_min)} (i.e. $f(x^*)$):\hfill\isalink{https://github.com/brydustin/Verification_Of_One_Dim_Optimization/blob/036eea319e62b5d65d809fcf9ca2defd9f312f80/Numerical_Methods/Gradient_Descent.thy\#L1178}

\begin{lstlisting}[style=isabellestyle,language=Isabelle,
  caption={Linear conv. bound for gradient descent under $L$-smoothness and PL},
  label={lst:gd-linear-conv}]
theorem gradient_descent_linear_convergence:
  assumes "0 < ~ε~" "0 < L" "0 < ~μ~" "~μ~ < L" "~⋀~x. ~α~ x = 1 / L" 
   and "~⋀~x. f x_min ~≤~ f x" "~⋀~x. (f has_derivative (~λ~h. grad x ~∙~ h)) (at x)"
   and grad_Lipschitz: "~⋀~x y. ~∥~grad x - grad y~∥~ ~≤~ L*~∥~x - y~∥~"
   and Polyak_Lojasiewicz: "~⋀~x. 2*~μ~*(f x - f x_min) ~≤~ grad x ~∙~ grad x"
  shows "H[True] gradient_descent (grad, x0, ~α~, ~ε~)
  [~∥~grad x~∥~~≤~~ε~ ~∧~ iter ~≤~ nat ~⌈~ln ((2*L*(f x0 - f x_min))/~ε~$^2$)/ln(1/(1- ~μ~/L))~⌉~]"
\end{lstlisting}

Polyak proved the original statement of the above formalisation in 1963~\cite{Polyak1963GradientMethods}, but our formalisation follows Theorem 1 of Karimi \textit{et al}~\cite{KarimiNutiniSchmidt2016PL}. However, we emphasise the bound on the number of iterations.  Critical to this development in Isabelle, we showed that $\nabla f$ being $L-$Lipschitz implies a well-known inequality in optimisation theory, namely,
$$f(y) \le f(x) + \nabla f(x)\bullet (y-x) + \frac{L}{2}\|y-x\|^2,$$ 
\noindent which in turn implies $\|\nabla f(x)\|^2 \le 2L \left(f(x) - f(x^*)\right)$. To assist in this verification, we demonstrated 4 corollaries about the Lebesgue integral defined in the file \texttt{Interval\_Integral.thy}~\cite{AvigadHoelzlSerafin_IntervalIntegral_Isabelle}. As customary, we present the invariants for proving the convergence of this method.

\begin{lstlisting}[style=isabellestyle,language=Isabelle,
  caption={Loop invariant and variant for total correctness of gradient descent under $L$-smoothness and the PL inequality},
  label={lst:gd-linear-conv-inv}]
invariant f(x)-f(x_min) ~≤~ (f(x0)-f(x_min))*(1-(~α~ x0)*~μ~)^iter
  ~∧~ iter ~≤~ nat~⌈~ln((2*L*(f x0 - f x_min))/~ε~$^2$)/ln(1/(1 - ~α~(x)*~μ~))~⌉~
  ~∧~ (~∥~grad(x)~∥~ > ~ε~ ~⟶~ 
  iter+1 ~≤~ nat~⌈~ln((2*L*(f x0 - f x_min))/~ε~$^2$)/ln(1/(1 -~α~(x)*~μ~))~⌉~)
variant nat~⌈~ln((2*L*(f x0 - f x_min))/~ε~$^2$)/ln(1/(1-~α~(x)*~μ~))~⌉~-iter
\end{lstlisting}

The invariant $\texttt{f}(x)-\texttt{f}(x_{\min}) \le (\texttt{f}(x_0)-\texttt{f}(x_{\min}))\cdot(1-(~\alpha~\,x_0)\,~\mu~)^{\mathrm{iter}}$ is precisely the convergence rate of Theorem 1 by Karimi \textit{et al.}~\cite{KarimiNutiniSchmidt2016PL} provided $\alpha(x_0)=1/L$. The variant and the second invariant match the postcondition in the total correctness triple, similar to the third invariant that also emphasises that if $\|\texttt{grad}(x)\| > \varepsilon$ then the current iterate cannot be equal to the upper bound. Again we use an auxiliary program to supply $L$ and $\mu$ for the proofs.

The third and strongest regime specialises $f$ to a strongly convex quadratic, $f(x)=\tfrac12\,x^\top Q x - b^\top x$ with $Q$ symmetric positive-definite, and replaces the constant step by \emph{exact line search}, $\alpha(x)=\dfrac{\nabla f(x)\bullet\nabla f(x)}{\nabla f(x)\bullet (Q\,\nabla f(x))}$. Writing $0<a\le A$ for bounds on the eigenvalues of $Q$ (obtained from an orthogonal diagonalisation $Q=U^\top D\,U$), the classical Kantorovich inequality yields the geometric per-step rate $\bigl(\tfrac{A-a}{A+a}\bigr)^2$, from which we again derive a closed-form iteration bound.\hfill\isalink{https://github.com/brydustin/Verification_Of_One_Dim_Optimization/blob/036eea319e62b5d65d809fcf9ca2defd9f312f80/Numerical_Methods/Gradient_Descent.thy\#L1859}

\begin{lstlisting}[style=isabellestyle,language=Isabelle,
  caption={Convergence at the Kantorovich rate for a strongly convex quadratic with exact line search},
  label={lst:gd-quadratic-conv}]
theorem gradient_descent_quadratic:
  assumes "0 < ~ε~" "0 < a" "a < A"
   and "~⋀~x. ~α~ x = (grad x ~∙~ grad x) / (grad x ~∙~ (Q *v grad x))"
   and "~⋀~x. f x = (1/2) * (x ~∙~ (Q *v x)) - b ~∙~ x"
   and "~⋀~x. (f has_derivative (~λ~h. grad x ~∙~ h)) (at x)"
   and "Q = transpose U ** D ** U" "orthogonal_matrix U" "diagonal_mat D"
   and "~⋀~i. a ~≤~ D ~\$~ i ~\$~ i ~∧~ D ~\$~ i ~\$~ i ~≤~ A" "D ** Dinv = mat 1" "grad x_star = 0"
  shows "H[True] gradient_descent (grad, x0, ~α~, ~ε~)
  [~∥~grad x~∥~ ~≤~ ~ε~ ~∧~ iter ~≤~ nat ~⌈~ln (A*((x0 - x_star) ~∙~ (Q *v (x0 - x_star)))/~ε~$^2$)/ln ((A+a)$^2$/(A-a)$^2$)~⌉~]"
\end{lstlisting}

Together, these three regimes already illustrate the spectrum of convergence criteria the method admits~\cite{KarimiNutiniSchmidt2016PL}; a fuller exploration is beyond the scope of this evaluation of the Isabelle/ITrees framework. We report proof effort for the linear regime as representative: proving linear convergence of \texttt{gradient\_descent} with \texttt{vcg} produces $12$ goals, $9$ of which can be discharged by \textsf{Sledgehammer}. The remaining $3$ require using our optimisation inequality from Lipschitz continuity and specifying various algebraic manipulations. Besides this, the results from this section rely on several auxiliary facts from elementary real analysis and integration. These include the evaluation $\int_a^b t,dt = (b^2-a^2)/2$, the monotonicity of the integral, namely that $f \le g$ on $[a,b]$ implies $\int_a^b f,dt \le \int_a^b g,dt$, a chain-rule lemma for composing a function with the line segment $t \mapsto x + t \cdot d$, and a specialised form of the Fundamental Theorem of Calculus on $[0,1]$.


\section{Evaluation}
\label{sec:eval}

Here, we track automation of proof obligations generated by \texttt{vcg} as a proxy for proof effort with the Isabelle/ITrees framework on the formalised numerical methods (assuming the library dependencies are available). We record the number of verification conditions, the percentage of those discharged by direct automation, and the lines of code (LOC) needed for each method in Table~\ref{tab:algorithm-eval}. The automation column counts VCs closed by a direct prover call, such as a \textsf{Sledgehammer}-suggested \texttt{by metis}, \texttt{by smt}, or a one-line simplification. It does not count obligations that need a structured Isar argument, even when that argument contains automated intermediate steps. LOC counts non-comment, non-blank Isabelle lines, excluding \texttt{text}, \texttt{section}, and \texttt{subsection} commands.

\begin{table}[t]
\centering
\caption{Verification statistics for the current Isabelle sources. The VCG counts were checked in Isabelle2025 against the built \texttt{Numerical\_Methods} session.}
\label{tab:algorithm-eval}

\vspace{-1ex}

\begin{tabular}{|l|c|c|c|}
    \hline
    \textbf{Algorithm} & \textbf{VCs} & \textbf{Auto.} & \textbf{LOC} \\ \hline
    Bisection & 15 & 13 (87\%) & 195\\
    FPM (local/quad.) & 4/4 & 1 (25\%)/2 (50\%) & 1111\\
    Perceptron & 11 & 8 (73\%) & 208\\
    Gradient descent (gen./PL/quad.) & 8/12/8 & 6 (75\%)/9 (75\%)/4 (50\%) & 1085\\ \hline
\end{tabular}

\vspace{-2ex}
\end{table}

The automation rates should be read together with the structure of the generated goals. Bisection is the most direct example: 13 of the 15 VCs close automatically, and the remaining two are symmetric root-extraction obligations that call the shared Bolzano helper. The perceptron is similar once the Novikoff invariants are supplied. Of its 11 VCs, 8 close directly; the remaining goals invoke the shared margin, norm-growth, and mistake-bound lemmas. In both cases, the difficult part is not low-level proof search but finding invariants and variants that expose the textbook argument in a form the VCG can use.

The FPM shows the other extreme. In the local contraction theorem, only one of four VCs closes directly. The other obligations are the convergence argument: geometric decay of the step size, the contradiction proof for the logarithmic iteration bound, and the error decay towards the fixed point. The quadratic theorem again generates four VCs, two of which are routine bound/variant obligations. The others use the Taylor--Peano expansion and the local contraction reasoning. This explains why the FPM LOC count is high: the imperative program is small, but the proof depends on substantial analysis infrastructure, including the separate higher-order differentiability developments not in the table.

Gradient descent lies between the two. The general smooth and PL-linear theorems both have 75\% direct automation: the guard-contradiction and exit-condition VCs are routine, while the remaining obligations apply the descent lemma, the PL contraction, or the iteration-bound lemmas. The quadratic exact-line-search proof has lower direct automation because the proof explicitly applies the Kantorovich contraction. Overall, \texttt{vcg} is effective at separating program reasoning from mathematical reasoning: it removes routine Hoare-logic bookkeeping, but the analytic content still has to be packaged as reusable Isabelle lemmas.

In general, proof effort tracks the mathematical content of the convergence theorem more closely than the size of the imperative program. Across the seven VCG invocations in Table~\ref{tab:algorithm-eval}, 43 of 62 obligations close directly, an aggregate rate of 69\%. The remaining goals correspond to recognisable textbook steps: root extraction, Novikoff estimates, local-contraction bounds, descent lemmas, and Kantorovich contraction. This supports the observation that Isabelle/ITrees gives an accessible front end for numerical programs, while scalability depends on reusable analysis and optimisation lemmas.

\section{Related Work}
\label{sec:related}

The literature on verification of numerical methods is abundant. In PVS, for instance, there has been work towards building libraries for reasoning about real numbers since at least the mid-2000s~\cite{DaumasLM09,GottliebsenHLM13,NarkawiczM14,ML05TPHOLs}. 
However, those works have focused on interval arithmetic and/or Taylor models as abstractions for reasoning about polynomial and real arithmetic inequalities. In contrast, we leverage the recent development of analysis libraries and directly reason about Taylor polynomials, rather than models. 

The Rocq community has produced vast work. From the formalisation of foundational theorems~\cite{TekriwalDJ2021} to the verification of specific numerical methods using similar methods of interval arithmetic, Taylor models and linear algebra~\cite{Pasca10} or complex combinations of Frama-C, automated provers and manual proofs~\cite{BoldoCFMM2012}. These approaches have converged into an actively developed framework for verifying numerical methods and producing reliable C programs~\cite{KellisonA22,AppelK24}. Just like our work, those projects leverage the libraries of a long-standing prover. For instance, it relies on the \texttt{Taylor\_Lagrange} theorem from the Coquelicot library~\cite{BoldoLM2015}. The framework is exhaustive, with a complete pipeline that accounts for numeric stability and rounding errors, including sound examples on finding the solutions to initial value problems of harmonic oscillators. In contrast, we have only focused on adapting an imperative programming framework into a method for verifying numerical methods. The \lstinline{execute} command only performs arbitrary precision operations. We consider the refinement of the Isabelle/ITrees framework to support numeric stability and rounding errors a worthwhile endeavour for future work.
Another, younger method in Rocq uses Taylor models and defines higher-order differentiation for approximating solutions to initial value problems for non-linear polynomial ordinary differential equations (ODEs)~\cite{ParkT2024}. It arbitrarily approximates the solution to the ODE by using more terms of its Taylor model.


In Isabelle/HOL itself, there have been verifications for specific algorithms, but not a framework per se. For instance, some Runge-Kutta methods over deep embeddings of arithmetic expressions or set-based Euler methods have been verified~\cite{Immler14,Immler18}. We have generalised the higher-order differentiability of the Smooth Manifolds AFP entry~\cite{ImmlerZ19} with our derivative definitions and shown their relationships. Furthermore, as the name suggests, the focus there is on formalising smooth manifolds instead of verifying numerical algorithms. The Isabelle/ITrees framework also has much in common with Lammich's Imperative HOL framework~\cite{lammich2019refinement}. However, we have endeavoured to avoid Isabelle idiosyncrasies in the program notation and VCG methods. We believe that such a choice lowers the barrier to using the Isabelle/ITrees framework, and means it can be applied without extensive knowledge of Isabelle. More recently, the pioneering work from Brucker and Stell~\cite{Brucker2023DNNsIsabelle} demonstrated the feasibility of verifying neural networks in Isabelle/HOL. The results presented here are complementary in that our work could pave the way for verifying the training algorithms, rather than the neural networks themselves.

Finally, in Lean, the rapid growth of the MathLib library~\cite{mathlib2020} has caught up with older provers like Rocq and Isabelle in terms of formalisation of higher-order derivatives, Taylor series and continuously differentiable functions. So much so that there is already work on constructing arbitrary stochastic computation graphs, and a proof of correctness for an executable stochastic backpropagation function~\cite{SelsamLD2017}. Yet, as far as we know, a verification framework for numerical algorithms is still nonexistent there.

\section{Conclusion}
\label{sec:concl}

We have evaluated Isabelle/ITrees, an imperative verification framework embedded in Isabelle/HOL, as a method for proving total correctness of numerical algorithms. The bisection method, fixed-point method, perceptron, and gradient descent all admit readable imperative programs, Hoare-style specifications, and loop annotations close to the textbook arguments. In each case, \texttt{vcg} reduces the proof to mathematical verification conditions; after suitable invariants and variants are supplied, many of these obligations are discharged by Isabelle's existing automation.

The case studies also make clear that the main cost is not programming the algorithms, but supplying the surrounding analysis. Bisection and the perceptron need little beyond their classical proof principles. Fixed-point iteration, by contrast, required higher-order differentiability and Taylor's theorem with Peano remainder to obtain sharp local and quadratic convergence statements, and gradient descent required reusable optimisation and integration lemmas. The accompanying Isabelle theories now also contain floating-point variants and Refinement Framework counterparts, which strengthens the evidence that the approach is not limited to idealised real-arithmetic models. Overall, Isabelle/ITrees provides a usable front end for verified numerical methods, while the scalability of the approach depends on continuing to package the analytic, optimisation, and floating-point facts that such methods routinely assume.

\paragraph{Funding statement:} A Horizon MSCA 2022 Postdoctoral Fellowship (project acronym DeepIsaHOL and number 101102608) supported the second author during the development of this article. The third author is partly funded by Horizon EU project \href{https://robosapiens-eu.tech/}{RoboSAPIENS} (grant agreement 101133807).

\paragraph{AI disclosure:} The structure of \texttt{descent\_lemma\_from\_Lipschitz\_gradient} was inspired from the output of ChatGPT 5.2 extended thinking. Claude Code with Opus 4.8 was used to generate the programs and proofs for the Lammich paradigm and floating point proofs.  No other Isabelle code in the contributions of this article has been produced with generative AI. 

\newpage

\bibliography{main}

\end{document}